# An Effective Early Multi-core System Shared Cache Design Method Based on Reuse-distance Analysis

Hsin-Yu Ho
National Tsing Hua University, Dept. of Compute Science
Hsinchu, Taiwan
shinyu951@gmail.com

Ren-Song Tsay
National Tsing Hua University, Dept. of Compute Science
Hsinchu, Taiwan
rstsay@cs.nthu.edu.tw

## ABSTRACT
In this paper, we proposed an effective and efficient multi-core shared-cache design optimization approach based on reuse-distance analysis of the data traces of target applications. Since data traces are independent of system hardware architectures, a designer can easily compute the best cache design at early system design phase using our approach. We devise a very efficient and yet accurate method to derive the aggregated reuse-distance histograms of concurrent applications for accurate cache performance analysis and optimization. Essentially, the actual shared-cache contention results of concurrent applications are embedded in the aggregated reuse-distance histograms and therefore the approach is very effective. The experimental results show that the average error rate of shared-cache miss-count estimations of our approach is less than 2.4%. Using a simple scanning search method, one can easily determine the true optimal cache configurations at early system design phase.

## 1. Introduction
A good cache design is known to be critical for system performance optimization. In the past, designers need to execute extensive simulations painstakingly to evaluate designs in order to achieve design objective of cost, performance and power consumption. Nevertheless, whenever a design configuration is changed, this time-consuming verification process has to be repeated endlessly. Recently, Tsai et al. have proposed a reuse-distance-histogram-based approach for accurate single core multi-level cache performance estimation and henceforth provided an effective early system-level cache design method [24]. Tsai showed encouraging experimental results with less than 4% average error rate of performance estimation. We in this paper further extend Tsai's effective approach for multi-core shared cache system designs.

Unlike the single core systems, certain caches are shared in the multi-core systems. Shown in Figure 1(a) is a typical 2-level-cache single core system with a first-level cache and a second-level cache, which also serves as the last level cache (LLC) before accessing main memory. Since only one application can take on a core at one time, the running application fully utilizes the entire last level cache space. However, for the two-core system illustrated in Figure 1(b), each core has a private first-level cache but shares the last level cache. Due to the cache sharing effect, the LLC hit rate of an application concurrently executing on a multi-core system in general is worse than the case if the application is solely using the LLC as in the single-core system.

In general for multi-core systems, different cores may share one cache and execute multiple applications concurrently. With a shared cache, an application often experiences performance degradation as compared to the case that the application solely uses the cache. The slowdown is mainly contributed by competing shared cache usage among applications. Allegedly, this resource sharing effect of concurrent applications has greatly increase design complexity. Clearly, it is critical to understand how the sharing mechanism affects system performance and is important to provide a systematic early system-level approach for multi-core designs. To ease discussions later in this paper, we adopt a two-level cache design for illustration. However, the conclusion can be generalized to general multi-level cache designs.

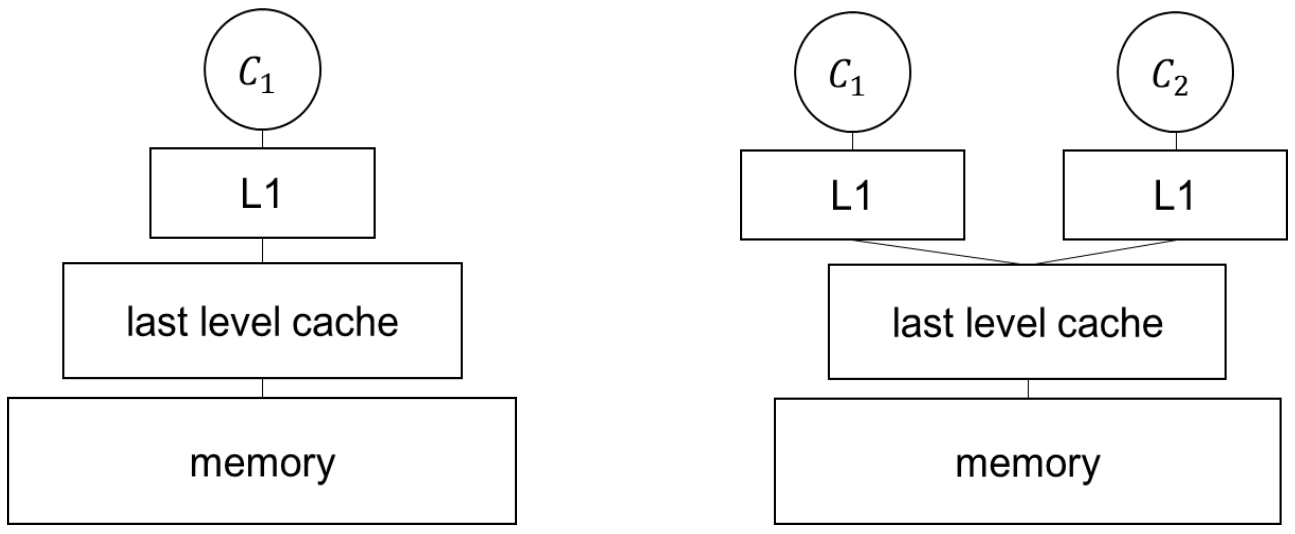

(a) (b)

Figure 1: A typical 2-level-cache processor architecture. (a) A single core system; (b) A two-core system.

To estimate hit/miss rate of a single-core application accurately on a one-level cache, Mattson et al. [16] has proposed using the reuse-distance histogram of the application. Each data access is associated with a reuse-distance number, which is defined as the number of accesses of distinct data memory addresses between current data access and the last access to the same data accessing

| | | | | | | | | | | |
|---|---|---|---|---|---|---|---|---|---|---|
| Memory trace | A | B | C | A | D | B | B | A | A | B | … |
| Reuse distance | ∞ | ∞ | ∞ | 2 | ∞ | 3 | 0 | 2 | 0 | 1 | … |

(a)

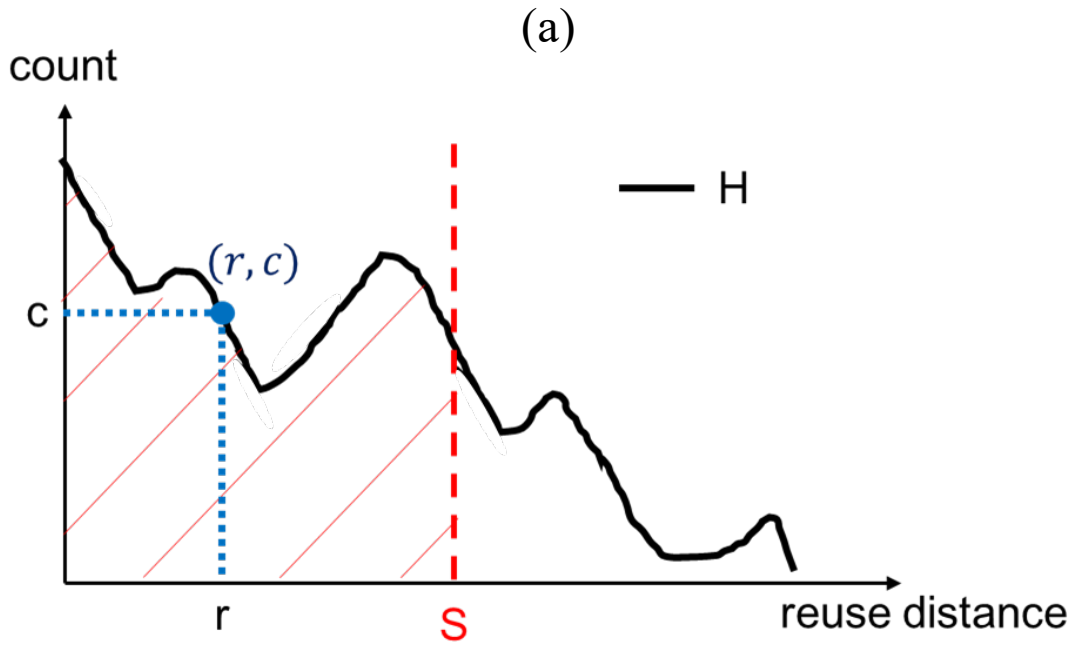

(b)

Figure 2: (a) A sample reuse-distance table. (b) An illustrative reuse-distance histogram example - H.

address. A data trace, or a series of memory addresses of continuous data access operations, is illustrated in Figure 2(a), in which the characters A, B, C … symbolize the addresses of data accesses, whereas the number under each access address is the corresponding reuse-distance number. Note that for the first-time occurring data address, the reuse-distance is always marked as ∞ (i.e. infinity), since there is no previous same address data access. Then, in between the $8^{th}$ memory access address, i.e. A, and the last same address access, at the $4^{th}$ data access, there are two distinct data access addresses, B and D. Therefore, the reuse-distance of the $8^{th}$ data access is 2. After collecting the reuse-distance count of each data access, we summarize the number of occurrences of each reuse-distance number and remake a reuse-distance histogram *(H)* as shown in Figure 2(b), with the reuse-distance number *(r)* on the horizontal axis and occurrence count *(c)* on the vertical axis. For instance, for the memory trace listed in Fig. 2(a), we have the reuse distance 0 occurs two times, reuse distance 1 occurs one time, and so on. In other words, a histogram *H* is the collection of all pairs of reuse-distance number and the corresponding occurrence count.

Mattson et al. [16] suggested that for any single-core design with fully associativity cache using LRU replacement policy, one could easily compute the first-level cache hit and miss counts using the reuse-distance histogram. For hit count, one simply sums up the occurrence numbers to the left of the reuse-distance number equivalent to the cache size, *S*, as illustrated by the red slashed area and the red vertical line in Figure 2(b). The corresponding miss count is the sum of the remaining occurrence numbers.

Tsai et al. [24] extended Mattson's method to compute the hit and miss counts accurately on multi-level caches requiring only the first-level reused-distance histogram, i.e. the one from CPU core to the first-level cache. In contrast, Mattson needs to use separate data histogram of each level to calculate hit rate. Considering inclusivity or exclusivity, Tsai accurately calculated hit/miss count of each level cache on the same reuse-distance histogram. A unique advantage of Tsai's approach is that once the target application is determined one can easily decide data trace and reuse-distance histogram independent of actual cache architecture. In other words, the data property analysis is decoupled from the hardware configurations. Therefore, a designer can easily compute the best cache design using the reuse-distance information of the target applications. Therefore, the method is perfectly suitable for early system-level designs. Encouraged by the success of this simple yet effective approach, we aspire to generalize the method to shared-cache designs.

Nevertheless, multi-core shared cache designs are much more complicated as multiple concurrently executing applications intefere with each other and perturb memory traces. For applications executed in a multi-core system, as illustrated in Figure 1(b), the memory trace consists of memory addresses from all cores in the system. The so derived reuse-distance histogram shall be named an *aggregated* reuse-distance histogram. Intuitively following the single-core case, we can still compute the multi-core hit and miss counts of concurrently executed applications from the aggregated reuse-distance histogram. Although one may indeed also determine the optimal cache size for sharing if having aggregated reuse-distance histograms, the issue is that one will need to generate the aggregated reuse-distance histograms of all possible application combinations in advance. Yet, it is very time-consuming and impractical to execute all application combinations.

Our major contribution in this paper is that we devise a very efficient and yet accurate method to derive the aggregated reuse-distance histograms from the single-core application reuse-distance histograms for accurate cache performance analysis and for optimum cache designs. Our experiments show that our proposed shared-cache hit/miss rate estimation method is highly accurate and efficient, the average error rate is less than 3.2%, and we can effectively design the optimum cache configurations. Details of our approach are to be discussed next.

This paper is organized as the following. We first review related work in Sec. 2 and elaborate the aggregated reuse-distance computation method in Sec. 3. Then, we summarize in Sec. 4 our experimental results and conclude the paper in Sec. 5.

## 2. Related Work

As a critical subject, many researchers have intensely studied how to improve shared cache usage for better system throughput. There are two general shared cache usage approaches: one tries to develop better shared-cache management methods for minimal cache contentions while another attempts to schedule best applications pair for concurrent executions. In general, these approaches all require accurate performance estimation methods based on pre-profiled or runtime information.

As for shared cache management for minimal access interference, some researchers developed shared cache partitioning methods for minimal performance degradation [6, 7, 8, 12, 13, 14, 15], while some proposed specific replacement policies for shared caches [1, 2, 3, 4, 5]. The cache partitioning methods limit each application using a certain part of shared cache space while the replacement policy approaches try to keep the frequently used data in cache. These approaches do help reducing the performance degradation but a general issue of these approaches is that they all require either additional hardware support or OS modification, which is a non-trivial task.

To avoid needing additional hardware support and OS modification but still be able to predict system performance, some researchers developed cache contention estimation methods based on certain application behavior information [9, 10, 11, 19, 20]. For example, Chandra et al. [9] and Xi E. Chen et al. [19, 20] profiled reuse-distance statistics (named circular sequences in their paper) of each thread from memory traces and devised probability models for shared cache miss rate prediction. In general, the probability models are very complex yet the estimation error rate is close to 10%. In contrast, our proposed reuse-distance histogram approach is much simpler while providing more accurate shared-cache miss rate calculations.

To avoid processing full memory traces, Eklov et al. [18] proposed a statistical cache sharing model based on certain partial (also called sparse) memory trace of each thread for aggregated reuse-distance estimation. Similarly, Xu et al. [10] and Sandberg et al. [11] predicted performance based on the shared cache access count derived from the performance counter of each application. Normally, these approaches apply a synthetic application that intensively accesses cache segments of various sizes and then construct the statistics of cache sizes and hit/miss rates. Given a cache access count, these methods estimate the effective cache size the target application occupies and the corresponding hit/miss rate under concurrent executions of other applications. In general, the issue of these approaches is that the partial information may not be representative. Additionally, it is not clear how the cache size or way number may affect the performance estimations. In contrast, our method can clearly reflect the effect of cache size or way number changes.

Instead of requiring pre-profiling, Subramanian et al. [17] performed a runtime method to estimate the performance impact due to shared cache contentions. Essentially, Subramanian observed that the performance of each application is proportional to the access rate of shared cache. They compared the measured shared cache access rate of an application under concurrent execution to the cache access rate when the application is executing alone. The ratio is then used to calculate the performance slowdown of concurrent executions. The issue of this approach is that it requires an additional auxiliary tag store hardware [6] to measure the access rate in runtime. Furthermore, the method is not applicable for early system designs.

In practice, instruction accesses also compete with data accesses and affect performance. Therefore, Jaleel et al. [25] observed that keeping the instruction cache lines in the shared cache could increase performance, but most other approaches did not consider the effect of instruction cache. With the aggregated reuse-distance approach, our approach take the instruction cache into account and can determine the best design configuration at early system design phase. Moreover, most existing approaches are limited to two-level cache designs whereas our approach effectively handles multi-level cache for multi-core systems.

Details of our proposed approach are elaborated below.

## 3. Shared Cache Design Optimization

As concluded in Tsai's research work [24], designers can now determine the optimal cache size for a single core system using reuse-distance histograms. We in this paper extend the idea by using aggregated reuse-distance histograms for optimization of the shared cache designs for multi-core systems. Essentially, an aggregated reuse-distance histogram describes how concurrently executed applications access the shared cache. Figure 3 illustrates how the memory traces of two concurrently executing applications are aggregated into one trace on a two-core system. The aggregated trace is then used to produce an aggregated reuse-distance histogram *H'* shown in Figure 4, with aggregated reuse-distance number *r'* on horizontal axis and occurrence count *c* on the vertical axis, similar to the reuse-distance histogram in Figure 2(b).

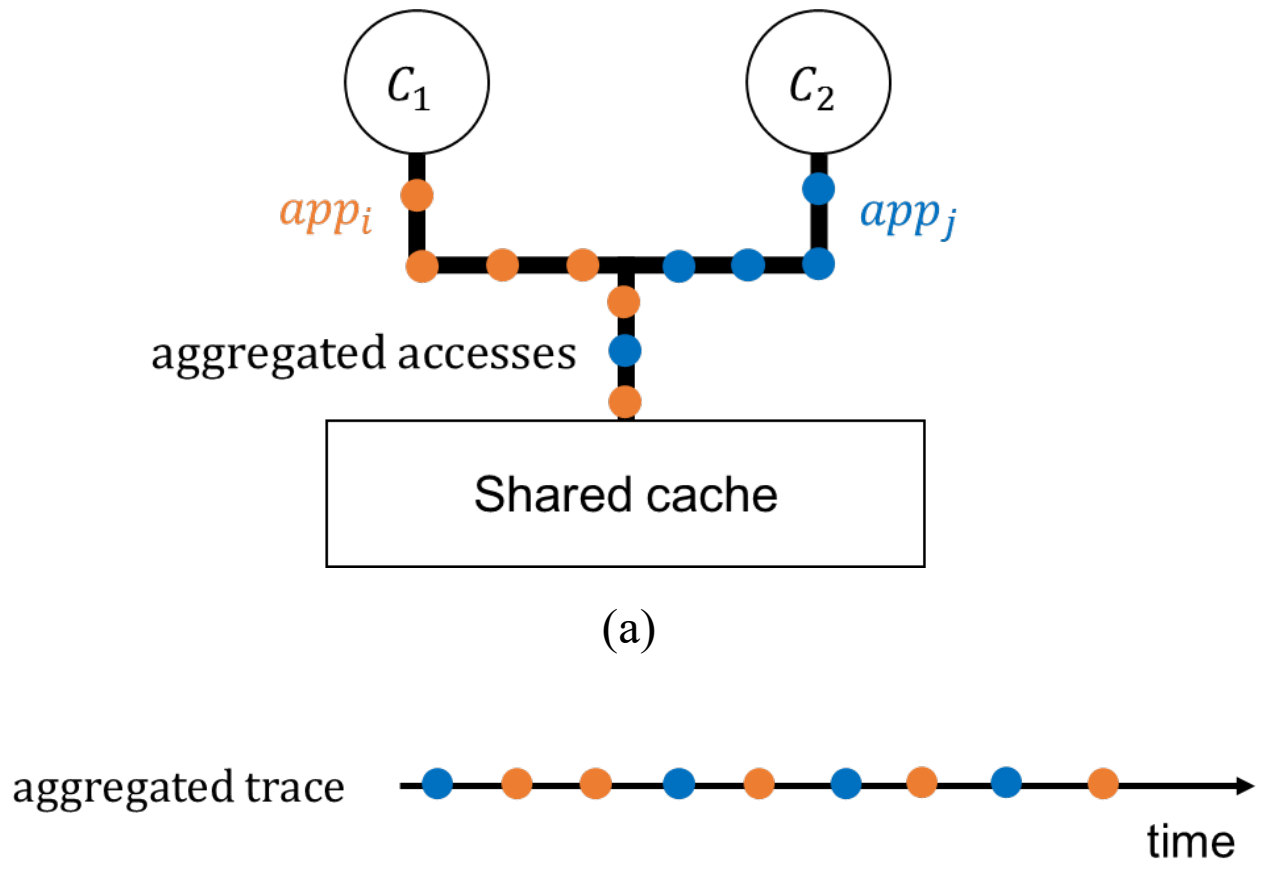

Figure 3: (a) An illustration of aggregated memory accesses from two concurrently executed applications. (b) A sample of an aggregated memory trace of the concurrent executed applications.

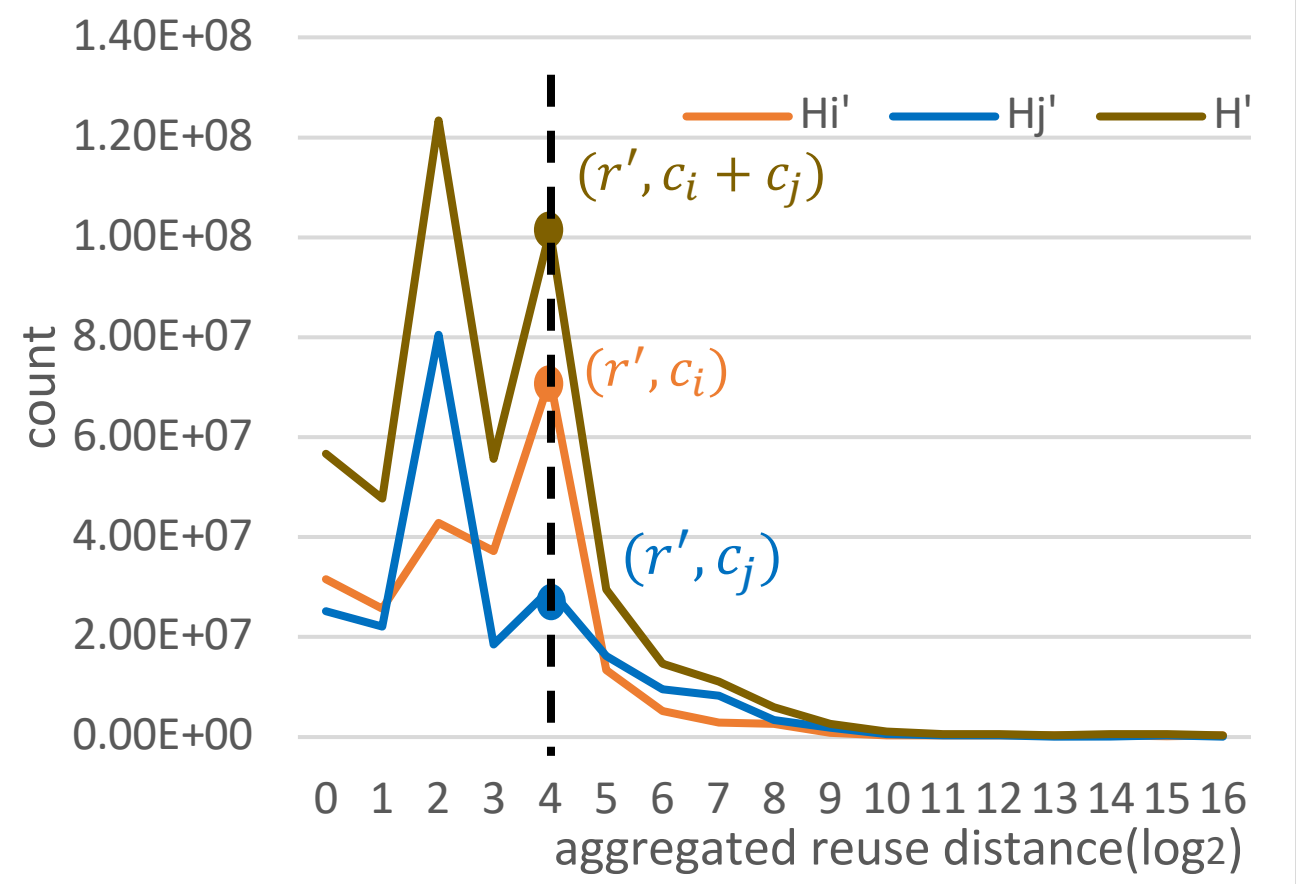

Figure 4: H' is the aggregated reuse-distance histogram of two concurrent executing applications $app_i$ and $app_j$. $H_i$' and $H_j$' are the corresponding aggregated reuse-distance histograms of $app_i$ and of $app_j$.

In practice, *H'* is normally produced from the aggregated reuse-distance histogram $H_i$' of $app_i$ and $H_j$' of $app_j$ as shown in Figure 4. Note that the aggregated reuse-distance for each individual application has to include other concurrent accesses. Then, for any particular reuse distance *r',* if the occurrence counts of $app_i$ and $app_j$ are $c_i$ and $c_j$ respectively, there shall be a corresponding data point $(r', c_i + c_j)$ on *H'*.

Now with the aggregated reuse-distance histogram *H'*, we may easily determine the optimal cache organization. Following the method of Tsai, once the cache size of each level is determined, we can effortlessly compute the cache miss rate of each cache hierarchical level from the aggregated reuse-distance histograms. Note that the cache miss rate of each application can be derived from $H_i$' and $H_j$'. In other words, with the aggregated reuse-distance histograms, we can precisely analyze the cache sharing effect of concurrently executing applications. However, in practice there are a few challenges we have to overcome.

First, although the aggregated reuse-distance histogram intuitively can be generated from the aggregated memory access trace of the concurrent executions of target applications as illustrated in Figure 3, running laborious simulation of every target application pair is very time consuming and impractical. For $n$ target applications, one will need to run simulations of $n(n-1)/2$ application pairs. For example, if 100 target applications are to execute on a two-core system, there are in total 4950 application pairs. Each simulation run can take days and therefore having complete simulations clearly is not feasible.

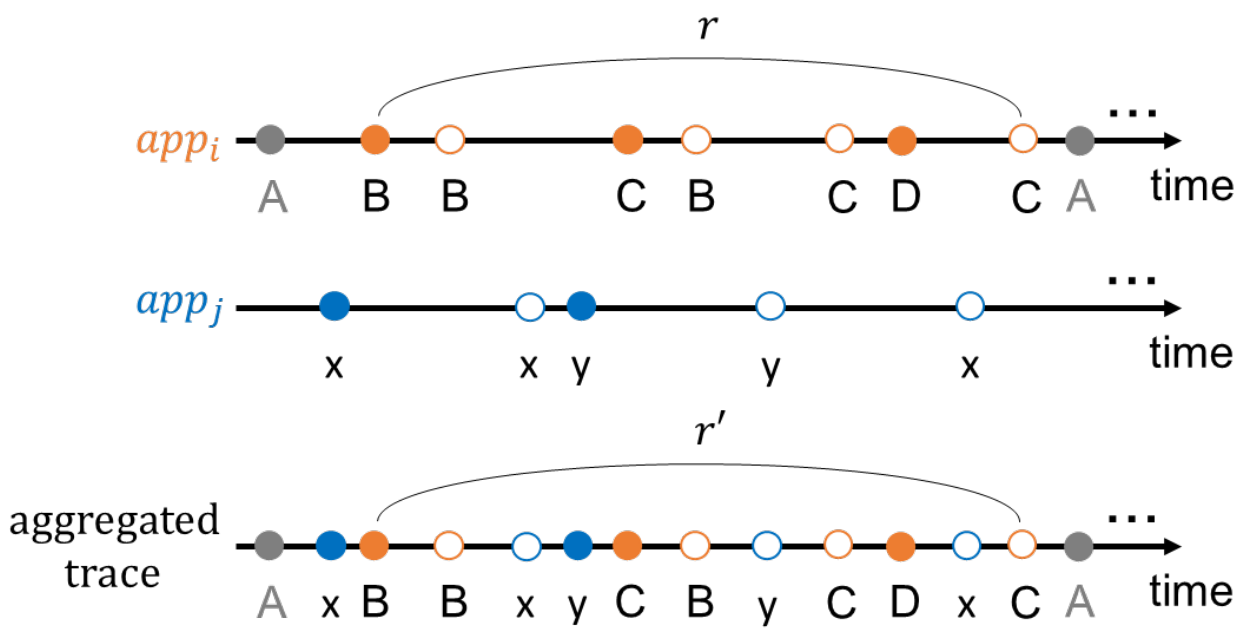

Figure 5: In the aggregated access sequence, the reuse-distance of a given memory access is to increase to a larger number due to access interleavings from the concurrent execution of other applications.

Furthermore, we observe a fact that with different cache size, the same application pair may produce different aggregated traces even if the CPU cores are fixed. This is because the cache size choice greatly affects cache miss behavior of an application and hence the merged trace sequence is changed accordingly, although the aggregated traces are in general very similar with common characteristics.

Therefore, to avoid the time-consuming concurrent pair simulations, we devise in the following an effective method that produces accurate aggregated reuse-distance histogram based on the memory trace of each individual application. The so generated aggregated reuse-distance histograms are verified through thorough experiments to be very accurate for performance analysis and cache design optimization. Details are elaborated in the next section.

### *3.1* Aggregated reuse-distance computation

We first observe that because of the memory access interleaving effect of concurrent application execution, the reuse distances of the data accesses of a stand-alone application shall in general increase in numbers in the aggregated memory accesses as illustrated in Figure 5. Suppose that the reuse distance of the second access to the address $A$ is $r=3$ for the stand-alone application $app_i$. The white dots in the figure represent repeatedly accessed addresses. With the memory access interleavings of a concurrent application $app_j$, the reuse distance of the same second access to $A$ shown in the aggregated trace of the concurrent execution is then increased to $r'=5$ as illustrated.

Specifically in Figure 5, we observe that although there are five data accesses from $app_j$ interleave in between the first and second accesses to the address A in the aggregated trace, there are only two new unique accesses, i.e. x and y. Therefore, the original reuse-distance number $r$ should be increased by two to become the new aggregated reuse-distance number, i.e. $r'=3+2=5$.

Our proposed idea is to develop an accurate method to calculate how the original reuse distance $r$ is to change to $r'$ in the aggregated access sequence. Essentially, we first compute how many data accesses from other concurrent applications are to come in between the current data access and the last access to the same address, and among these two accesses how many are unique accesses (accesses to unique addresses). If the original reuse distance is $r$ and the number of the new unique accesses is $u_r$, then the aggregated reuse distance is $r'=r+u_r$. In other words, if we can compute the number of unique ones among the interleaving accesses, we can compute the aggregated reuse distance $r'$ without need to generate the aggregated trace.

We now discuss how to compute the number of data accesses that may interleave from other concurrent applications and how to calculate the number of unique accesses among them so that we can compute the increase of the reuse distance count.

Although without aggregated trace we cannot directly compute the number of interleavings from other concurrent applications, we may compute the average interleaving accesses between any two consecutive data accesses of the application of concern. Assume that the application of concern is $app_i$ and the concurrent application is $app_j$. In addition, the numbers of data accesses of $app_i$ and $app_j$ are $n_i$ and $n_j$ respectively for a fixed time period of execution. Then the average number of $app_j$ data-access interleavings between two consecutive data accesses of $app_i$ is $n_j/n_i$, which is named the *access ratio* of $app_j$ to $app_i$.

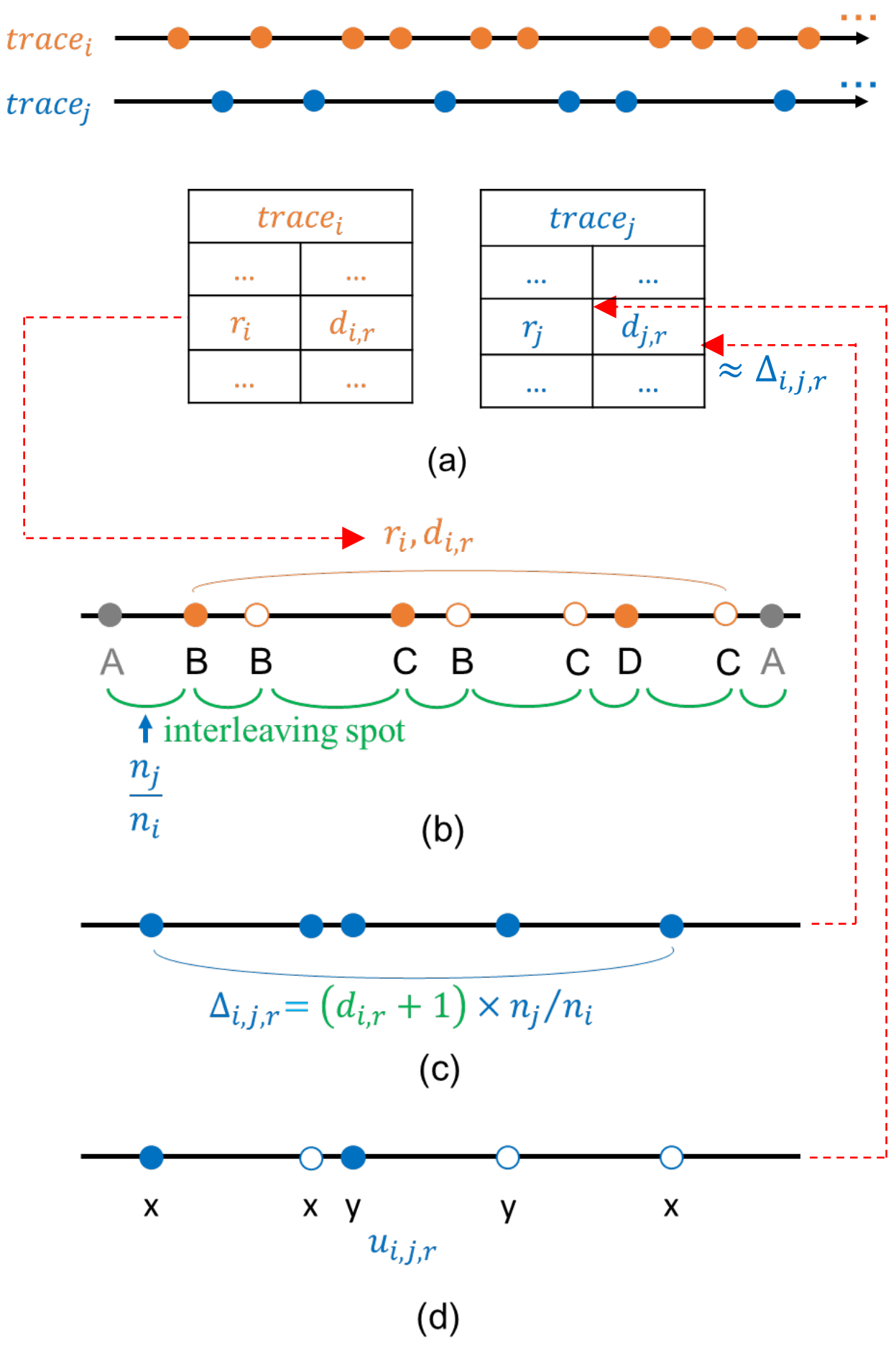

Figure 6: An illustration of the aggregated reuse-distance computation flow. (a) The *r-d* tables of application traces. (b) Find the average data-access number $d_{i,r}$ for each reuse-distance number $r_i$ from the r-d table; (c): Use the access ratio to compute the interleaving data accesses number $\Delta_{i,j,r}$ from $app_j$; (d) Use the r-d table to compute the unique access number $u_{i,j,r}$ from $\Delta_{i,j,r}$.

We illustrate in Figure 6 the aggregated reuse-distance computation flow, in which the orange dots and blue dots represent the data accesses from $app_i$ and $app_j$ respectively. Since only the data trace (and reuse-distance histogram) of each application is available, but not the aggregated trace, we first compute the average data-access number of the accesses correspond to each reuse distance number of the application. In general, for each application $app_i$, we associate each reuse distance $r_i$ as analyzed from the data trace an average data access number $d_{i,r}$, which is no less than $r_i$. We then

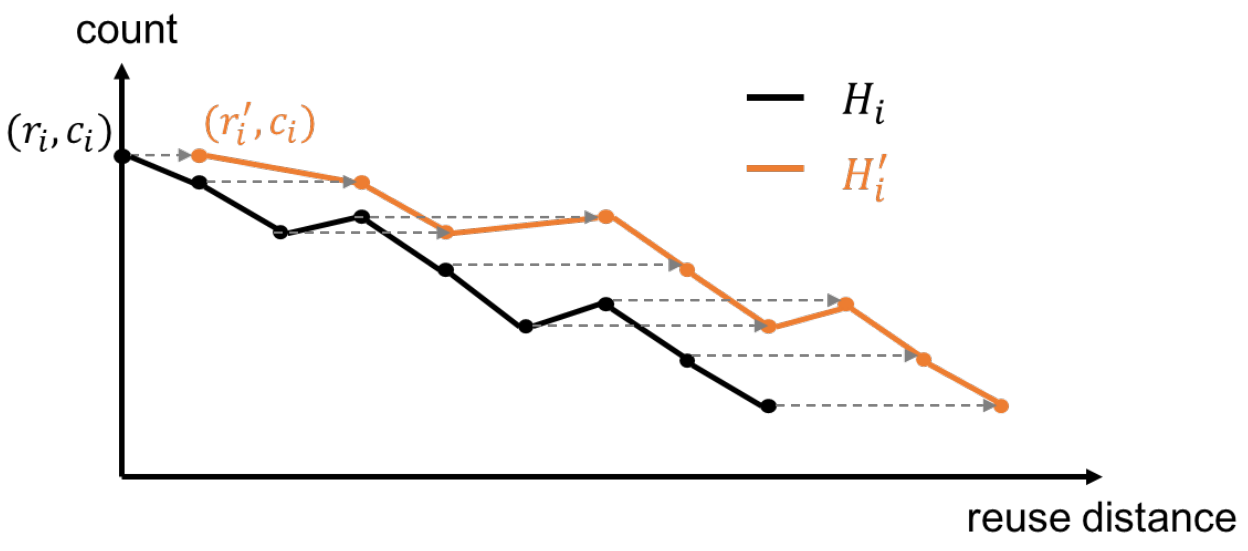

Figure 7: The aggregated reuse-distance histogram of $app_i$ affected by $app_j$ can be produced by shifting each original reuse-distance point ($r_i$, $c_i$) horizontally to ($r'_i = r_i + u_{i,j,r}$, $c_i$).

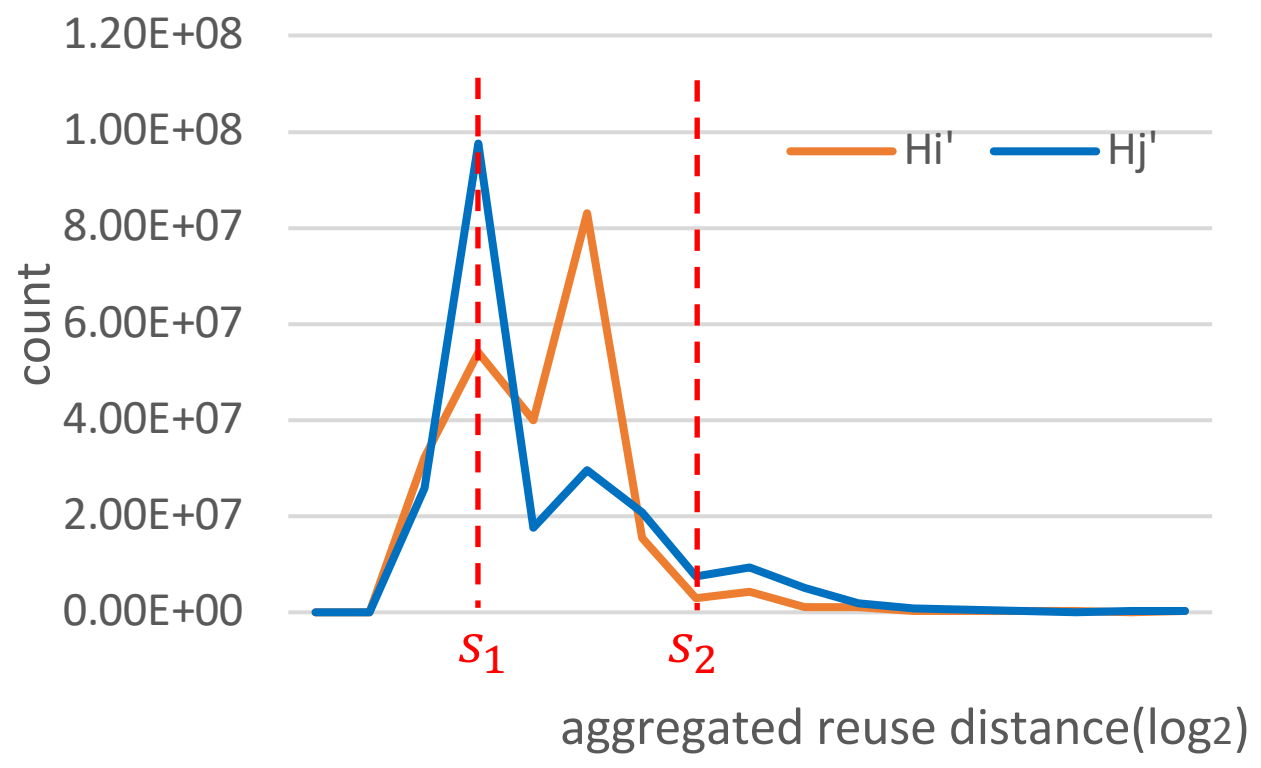

Figure 8: Given a cache size, one can compute the cache miss rate using the aggregated reuse-distance histograms of the concurrently executing applications

generate for each application an *r-d table* listing reuse-distance numbers and the corresponding average data-access numbers as shown in Figure 6(a).

Now for each reuse-distance number $r_i$ of $app_i$, we can look up the corresponding average data accesses number $d_{i,r}$ from the r-d table as illustrated in Figure 6(b).

Once we have the average data access number $d_{i,r}$, we use it to compute the average reuse-distance increase $u_{i,j,r}$ contributed by the data interleavings from $app_j$. In fact, along with the $d_{i,r}$ data accesses, there are ($d_{i,r}+1$) possible interleaving spots as illustrated in Figure 6(c). Consequently, in average there are $\Delta_{i,j,r}=(d_{i,r}+1)^* n_j/n_i$ interleaved data accesses from $app_j$ as shown in Figure 6(d), since each interleaving spot can have $n_j/n_i$ interleaved data accesses.

Supposedly, different applications access disjoint data sets. Therefore, the remaining task is to translate the increased number of the interleaved data accesses, $\Delta_{i,j,r}$, to an equivalent increased number of reuse-distance count $u_{i,j,r}$. This is the same as asking how many unique-address data accesses are there for $\Delta_{i,j,r}$ data accesses in $app_j$. In fact, this is an inverse calculation as that of computing the data access numbers from the reuse distance numbers.

With the average access number $\Delta_{i,j,r}$ from $app_j$, we find the $d_{j,r}$ which is closet to $\Delta_{i,j,r}$ from the r-d table of $app_j$. Then we use the corresponding reuse-distance numbers $r_j$ to represent the unique access number $u_{i,j,r}$, as illustrated in Figure 6(d).

Then with this calculated increased reuse distance number $u_{i,j,r}$, we shift, in the reuse-distance histogram $H_i$ of $app_i$, the original reuse distance point ($r_i$, $c_i$) horizontally to ($r'_i = r_i + u_{i,j,r}$, $c_i$) and have a new aggregated histogram $H_i'$, as illustrated in Figure 7. Similarly, we can compute the aggregated histogram $H_j'$ from the reuse-distance histogram $H_j$.

With the aggregated reuse-distance histograms $H_i'$ and $H_j'$ shown in Figure 8, then according to Tsai's work [24], one can accurately compute the shared cache hit and miss counts of $app_i$ and $app_j$ on multi-level caches. This method is verified to be very accurate by experiments as discussed in Section 4.

## *3.2* Cache configuration optimization

Our approach can be easily applied in early system design phase to determine the optimal cache configurations. In contrast, most existing works assumed that the cache configuration is pre-determined and focused on developing shared-cache partitioning or application-pair scheduling methods for minimal last-level cache miss rate, minimal slowdown or maximal throughput [15, 8, 12]. In fact, one can obtain better results by optimizing the cache configurations using our proposed approach as discussed below.

The proposed aggregated reuse-distance-histogram cache design approach is perfect for determining optimal cache configurations at the early-system design phase. The approach can be applied to any given objective function. For instance, one possible objective cost function $f$ for $n$-level cache is the total slowdown due to cache misses, i.e.

$$f(s_1, \dots, s_n) = \sum\nolimits_{l=1}^{n} d_l[M_{l,i}(s_l) + M_{l,j}(s_l)],$$

where $d_l$ is the delay penalty of the $l$-th level cache miss, $M_{l,k}(s)$ is the miss count of $app_k$ on the $l$-th level cache of size $s$. To calculate the miss count $M_{l,k}(s)$, following Mattson's method [16] one simply sums up the occurrence numbers to the right of (including) the reuse-distance number equivalent to the cache size $s$ on the derived aggregated reuse-distance histograms $H_k'$ as illustrated in Figure 8.

If we have $d_n=1$ and $d_{l\neq n}=0$, then the objective function

$$f = M_{n,i}(s_n) + M_{n,j}(s_n)$$

represents the total miss count of the last-level cache. The shared-cache partitioning methods [15] mainly try to minimize the total last-level cache miss count with given cache sizes.

Note that at the early system design phase, cache sizes are to be determined and not fixed. Normally we shall have a total cache cost limit, otherwise unlimited sized caches always give the minimum miss count. Assume that the total cache cost $g$ is

$$g(s_1, \dots, s_n) = \sum_{l=1}^{n} a_l s_l,$$

where $a_l$ is the per unit sized cache cost of the $l$-th level cache.

Then a formal $n$-level cache design optimization problem can be formulated as the following,

$$\begin{aligned} min: \quad & f(s_1, \dots, s_n) = \sum_{l=1}^{n} d_l[M_{l,i}(s_l) + M_{l,j}(s_l)] \\ s.t. \quad & g(s_1, \dots, s_n) = \sum_{l=1}^{n} a_l s_l \le G, \end{aligned}$$

where $G$ is a designer specified total cache cost limit.

Another possible formulation is to minimize total cache cost under a maximum slowdown limit as the following,

$$\begin{aligned} min: \quad & g(s_1, \dots, s_n) = \sum_{l=1}^{n} a_l s_l \\ s.t. \quad & f(s_1, \dots, s_n) = \sum_{l=1}^{n} d_l[M_{l,i}(s_l) + M_{l,j}(s_l)] \le F, \end{aligned}$$

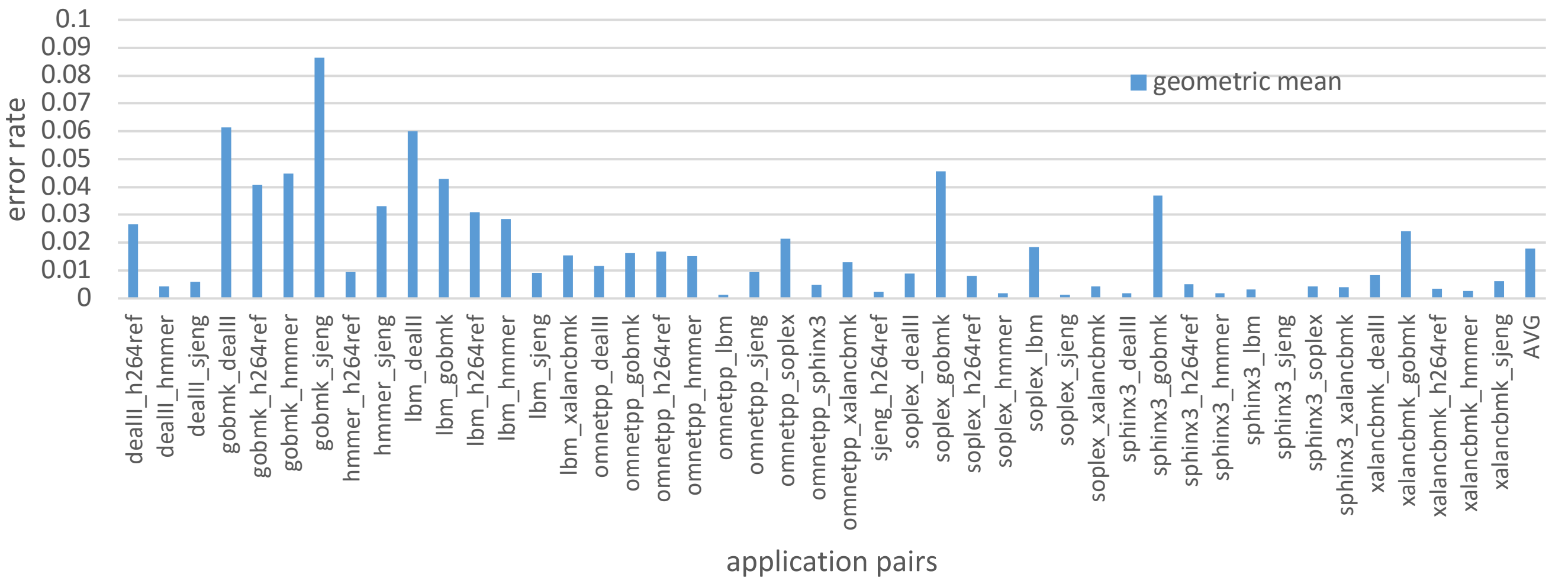

Figure 9: The average error rates of the miss count estimations of 45 application pairs for one-level shared cache architecture.

where *F* is a designer specified slowdown limit.

In order to find the optimal solution, we observe the fact that since the cache size is always in the form of $2^n$, it is of limited number of choices. As suggested by Tsai [24], we may then adopt the scanning approach to find the optimal solution. Essentially, we evaluate the cost function of each possible cache size used in practice and find the corresponding optimal cache configuration that produces the minimal cost within constraints.

In summary, our proposed approach is very flexible and can be applied to evaluate any designer specified cost and constraint functions. Particularly, the approach is effective for early multi-core system shared cache designs. Next, we show experimental results to support the effectiveness of our proposed approach.

# 4. Experimental Results

## *4.1* Experimental setup

We follow Jaleel et al. [24] and take the same ten representative test cases, from SPEC CPU2006 [23] suite, listed in Table 1 to test our proposed approach. The ten representative test cases include three high-CPI memory-bound cases, four low-CPI memory-bound cases, and three CPU-bound cases. In fact, there are in total 29 SPEC CPU2006 test cases and 406 pairs of applications. With ten representative test cases, we only need to verify 45 pairs of applications and save 10x testing time while still having solid, unbiased results.

Table 1: Benchmarks tested from SPEC CPU2006.

| Benchmark type | Benchmarks |
|---|---|
| **Memory bound(CPI=4-8)** | omnetpp, soplex, lbm |
| **Memory bound(CPI=2-4)** | gobmk, hmmer, sphinx3, xalancbmk |
| **CPU bound** | dealII, h264ref, sjeng |

The multi-core simulator Sniper [22] is used to compute the actual miss rates, cache and execution performance for comparison and validate the estimated results of our approach. In general, we execute the given pair of applications on Sniper, and have the faster application executes 400 million instructions. Along with the execution, we record the data access trace of each application and compute at the same time the cache miss counts for later references.

## *4.2* Results of One-level Shared Cache Designs

Before we test on more realistic two-level cache designs, we first conduct testing on an in-order two-core one-level cache architecture, having instructions come from memory directly and two cores share the one-level data cache. The shared cache is fully associative with LRU replacement policy. The purpose of this test is to validate the accuracy of the proposed aggregated reuse-distance histogram generation method.

We use the miss counts computed from Sniper simulations as references and calculate the error rate ε as the following,

$$\varepsilon = \frac{m_s - m_r}{m_s},$$

where $m_s$ is the cache miss count computed from Sniper simulation results and $m_r$ is the miss count estimated by our proposed aggregated reuse-distance computation model. We test ten applications listed in Table 1, which has 45 application pairs. The experiments show that the geometric mean of various cache sizes (512KB, 1MB and 2MB) is less than 1.8%, and detailed results are shown in Figure 9. The results show that our approach is accurate.

## *4.3* Results of Two-level Cache Designs

To be practical, we also conduct experiments on in-order two-core, inclusive two-level-cache designs with 64-Byte line sized caches. Each core owns a private L1 instruction and a data cache but shares an L2 cache, which is the last-level cache to the main memory. The access latencies of the L1, L2 caches and main memory are 1 cycle, 10 cycles and 130 cycles respectively. We also use the 45 application pairs as the test cases.

In addition to fully-associative cache, all experiments are conducted also on 8-way and 16-way set-associative caches to be practical, although theoretically the reuse-distance-based estimation of cache miss count matches perfectly only for fully associative, LRU-replacement-policy cache designs.

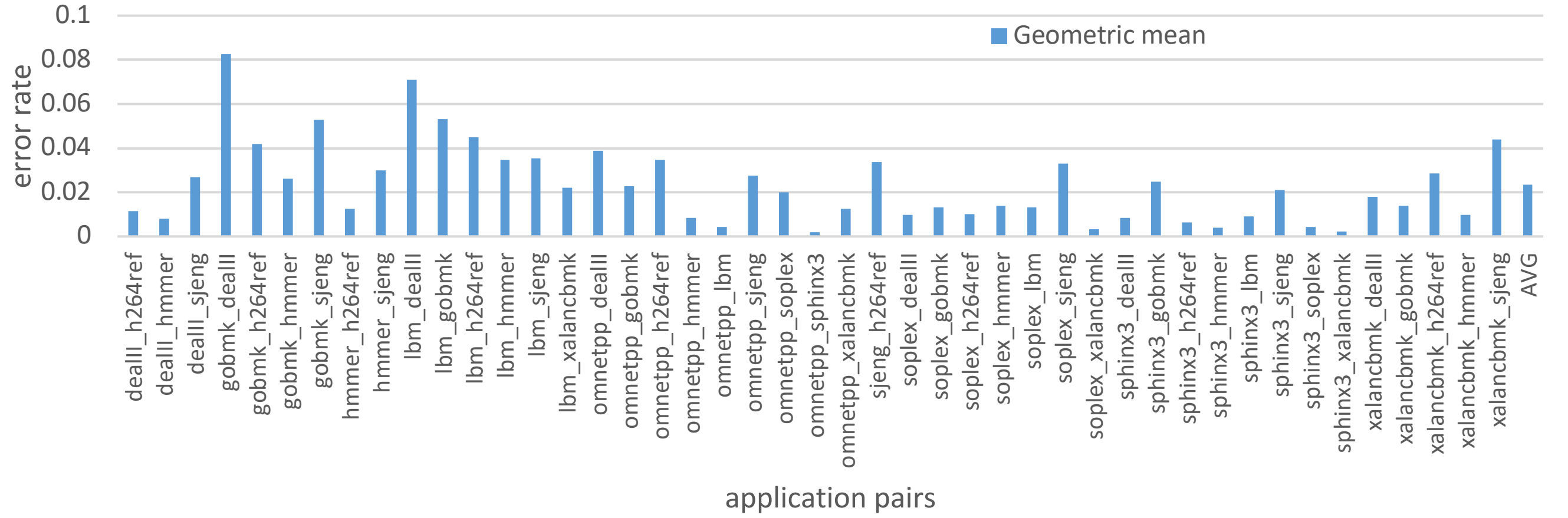

Figure 10: The geometric mean of the miss count estimations of 45 application pairs for a two-level shared cache architecture.

The cache configurations with various sizes and way-associativity are shown in Table 2. In Table 2, the two numbers $x_1_x_2$ on the first row indicate that the L1 (private instruction and data) cache is of size $x_1$ bytes and the shared L2 cache is of size $x_2$ bytes. Next, the two numbers $y_1_y_2$ on the second row indicate that the L1 cache is $y_1$-way set associative and the shared L2 cache is $y_2$-way set associative. The symbol $F$ indicates a fully associative cache.

Figure 10 shows the geometric mean of the shared L2 miss counts estimated by our approach and that calculated from Sniper simulations. In general, the total average error rate of 45 application pairs is less than 2.4%.

Table 2: Combinations of different cache size and way-associativity form various 2-level cache configurations.

| Cache size | 16KB_512KB | 32KB_1MB | 32KB_2MB |
|---|---|---|---|
| Way-associativity | 4_8 | 4_16 | F_F |

The extremely low error rates support the fact that our proposed aggregated reuse-distance-base computation model is very accurate for multi-core and multi-level-cache designs. Therefore, once the cache size of each level is determined at the early design stage, we can effortlessly compute the cache miss rate of each cache hierarchical level based on the aggregated reuse-distance histograms. Henceforth, we may optimize cache configuration for best system performance for any given target applications.

## *4.4* Optimal Cache Designs

To show that our proposed method can effectively determine optimal cache designs, we compare our results with that of the partitioning approach discussed in section 3.2. Assume that the objective function is simply to minimize the total miss count of the shared last-level cache, L2.

For comparison, we implement the partitioning method on a design with 8-KB private L1 cache and 1-MB L2 shared cache. We find that by varying L1 and L2 cache sizes, we always obtain better performance results than that of the partitioning method in terms of total execution cycle counts. The partitioning method focuses on minimizing the miss counts of L2, or equivalently the number of main memory accesses, in order to increase performance. In fact, the miss count of L1 also affects the performance, and our approach considers the total effect and hence get the best results. The access latencies of the L1, L2 caches and main memory are the same as that used in the last section, i.e. 1 cycle, 10 cycles and 130 cycles respectively.

Specifically, for the application pair *dealII* and *sjeng,* by increasing the L1 cache to 16 KB while decreasing the L2 cache to 512 KB, the total number of cycle count to complete the execution is 7.24% less than the best partitioning results on the original given cache sizes.

In general, the partitioning method requires hardware modification and has additional hardware cost. With the accurate aggregated reuse-distance computation model, we can now decide the optimal cache size of multi-level cache at the early design phase without requiring unnecessary cost.

## *4.5* Discussions

The error sources of our approach mainly are contributed by the following three factors. First, the major one is due to the estimation error of the proposed aggregated reuse-distance histogram generation method. However, the experiments show that the average error rate is only 2%. Actually, as discussed in the main body of the paper, we may conduct simulations of the target application pairs to get more accurate histogram data, but then we will need to assume certain cache sizes and take long simulations to obtain the results. This will then defeat the purpose of being able to perform early system-level designs and to determine the optimal cache sizes. Therefore, the estimation-induced error is a choice of our tradeoff decision.

Secondly, the invalidated LRU records on the higher-level inclusive cache also contributed to errors. Note that for an inclusive cache architecture, the data on an L1 cache always has a replica in L2. When a victim data of L1 is evicted and replaced, the replica is still in L2 with an outdated LRU value, which theoretically should be inherited from the corresponding L1 victim data, following most practiced cache implementations. The inconsistency contributes slight inaccuracy.

Finally, since real designs are mostly of k-way associativity, not the ideal fully associativity, the modeling discrepancy also causes errors. In Tsai's experiments [24], the results show that the above two factors contributes only negligible errors.

## 5. Conclusion

In this paper, we have proposed an accurate aggregated reuse-distance computation model to estimate the shared cache hit/miss counts without needs to run simulations. Using a simple scanning search approach, we efficiently and effectively explore multi-level cache configurations for optimization. Our proposed approach does not require hardware support or OS modification, and is perfect for early system design use. For future work, we may extend the idea to perform complete storage system optimization.